\documentclass[%
 reprint,
 amsmath,amssymb,
 aps,
]{revtex4-2}

\usepackage{graphicx}
\usepackage{dcolumn}
\usepackage[dvipsnames]{xcolor}
\usepackage{bm}
\usepackage[hidelinks]{hyperref}
\usepackage{float}
\usepackage{soul}
\usepackage{upgreek}

\begin{document}

\title{Imaging of Electrical Signals in a Quantum SiC Microscope}

\author{A.~Suhana$^{1,2}$} \email{a.suhana@hzdr.de}
\author{T. A. U. Svetikova$^1$}
\author{C. Schneider$^1$}
\author{M. Helm$^{1,2}$ }
\author{A. N. Anisimov$^1$} \email{a.anisimov@hzdr.de }
\author{G. V. Astakhov$^1$}\email{g.astakhov@hzdr.de }

\affiliation{%
 $^1$Helmholtz-Zentrum Dresden-Rossendorf, Bautzner 
 Landstra{\ss}e 400, 01328 Dresden, Germany
 \\ $^2$Technische Universit\"{a}t Dresden, 01062 Dresden, Germany}

\begin{abstract}
We report the experimental realization of a quantum silicon carbide microscope (QSiCM) and demonstrate its functionality by imaging magnetic fields generated by electrical currents. We employ a dual-frequency sensing protocol to enhance the readout contrast and suppress noise arising from strain and temperature fluctuations. This approach enables spatial imaging of current-induced magnetic fields with a field of view of $50 \times 50 $ virtual pixels, temporal resolution of $50\,\mathrm{ms}$, spatial resolution of $30\,\mathrm{\mu m}$ and sensitivity of about $2\,\mathrm{\mu T \, Hz^{-1/2}}$ per pixel. Further sensitivity enhancement is anticipated through the use of isotopically purified SiC and improved light collection in crystallographically optimized wafer orientations. In addition, we implement a microwave-free imaging protocol based on spin level anticrossing, offering simplified operation with enhanced sensitivity. The demonstrated platform is compatible with commercial, wafer-scale fabrication and holds strong potential for applications in biomedical imaging and diagnostics, as well as non-invasive current and temperature mapping in high-power electronic devices.

\end{abstract}

\date{\today}

\maketitle


\section{\label{sec:introduction}Introduction}

Optically addressable spin defects in wideband gap materials such as diamond, silicon carbide (SiC) and hexagonal boron nitride (hBN) are promising material platform for room-temperature quantum sensing \cite{10.1038/nature07278, 10.1038/nphys1075, 10.1038/srep05303, 10.1038/s41563-020-0619-6}.  
Among these platforms, the nitrogen-vacancy (NV) defect in diamond is the most advanced, offering the highest sensitivity \cite{10.1038/nphys3291, 10.1103/physrevapplied.21.064010}. In contrast to scanning quantum magnetometer \cite{10.1021/acs.nanolett.4c03044, 10.1038/s41467-025-55956-1}, a quantum diamond microscope (QDM) captures the entire magnetic image in a single shot. This allows for significantly faster measurements and enables the detection of dynamic processes either in real time or through triggered acquisition. Particularly, QDM has been used for widefield imaging of  single cells \cite{10.1038/nmeth.3449}, detection of single-neuron action potential \cite{10.1073/pnas.1601513113} and immunomagnetic microscopy of tumor tissues \cite{10.1073/pnas.2118876119}.  Apart from biomedicine \cite{10.1038/s42254-023-00558-3}, QDM is a powerful tool to study solid-state magnetism \cite{10.1063/5.0169521, 10.1021/acs.nanolett.8b03272},  investigate geological samples \cite{10.1002/2017GC006946, 10.1029/2020GC009147, 10.1142/S201032471740015X} and characterize spatial distribution of electrical current  in electronic devices \cite{10.1103/PhysRevApplied.14.014097, 10.1364/OE.514770, 10.1103/PhysRevApplied.20.014036}.

The boron vacancy  ($\mathrm{V_{B}}$) in hBN exhibits  lower sensitivity compared to the NV defects in diamond \cite{10.1038/s41467-021-24725-1, 10.1038/s41467-023-40473-w}. However, 2D nature of hBN provides a unique capability for sensing in close proximity to the target object \cite{10.1080/23746149.2023.2206049}, which is essential for \textit{in situ} microscopy of van der Waals heterostructures  \cite{10.1038/s41567-022-01815-5}.  

On the other hand, SiC possesses an advantage for quantum technologies as it is readily available at the wafer scale, owing to its widespread use in high-power electronics and electric vehicles. Among its quantum-relevant defects, the silicon vacancies ($\mathrm{V_{Si}}$)  has half-integer spin $S = 3/2$, which offers some advantage compared to integer spin systems \cite{10.1088/0268-1242/23/11/114001, 10.1038/srep05303, 10.1103/physrevx.6.031014, 10.1103/physrevb.95.081405, 10.1038/s41467-019-09873-9, 10.1038/s43246-025-00840-0}. These properties are particularly promising for integrated magnetometry and thermometry \cite{10.1038/srep05303, 10.1103/physrevapplied.4.014009, 10.1103/physrevapplied.6.034001, 10.1103/physrevx.6.031014, 10.1063/5.0027603, 10.1103/physrevapplied.15.064022, 10.1103/physrevapplied.19.044086, 10.1038/s41563-023-01477-5, 10.1103/physrevapplied.20.l031001, 10.1126/sciadv.abj5030, 10.1364/ol.476305, 10.1126/sciadv.adi2042, 10.1038/s41534-025-01011-2, 10.1103/pv13-vgcw, 10.1021/acs.nanolett.5c02515}, opening the door to widefield imaging of magnetic fields, temperature and electrical currents in both electronic devices and biological systems.
A prominent example is thermoelectric monitoring in high-power devices \cite{10.1063/5.0027603} and electric field mapping in semiconductor devices \cite{10.1103/pv13-vgcw}. 

The core element of a QDM is either a lock-in camera or a scientific complementary metal-oxide-semiconductor (sCMOS) camera \cite{10.1063/5.0142448, 10.1116/5.0230098, 10.1116/5.0176317, 10.1116/5.0222809}. Each pixel of the camera functions as an individual quantum sensor based on optically detected magnetic resonance (ODMR). It requires optical excitation and photoluminescence (PL) detection, while the PL intensity is modulated using resonance microwave (MW) fields. A quantum SiC microscope (QSiCM) can be realized based on the same principle; however, it has not been demonstrated to date. In this work, we present the implementation of a QSiCM using a commercial SiC wafer. Its functionality is demonstrated with three distinct sensing protocols, enabling imaging of magnetic fields generated by electrical pulses in a wire with spatial and temporal resolution.


\section{\label{sec:methods}Experimental Section}



A 4-inch high-purity semi-insulating (HPSI) silicon carbide (4H-SiC) wafer with a thickness of $300\,\mathrm{\mu m}$ was diced into $2 \,\mathrm{mm} \times 2 \,\mathrm{mm} $ and $2 \,\mathrm{mm} \times 5 \,\mathrm{mm} $ samples. The wafer is nominally undoped and has resistivity $\geq 1 \times 10^{7}\,\mathrm{\Omega \cdot cm}$ at room temperature. To create homogeneously distributed silicon vacancies ($\mathrm{V_{Si}}$) along the entire samples, we performed irradiation with $30\,\mathrm{MeV}$ electrons to a fluence of $2\times 10^{18}\ \text{cm}^{-2}$ at the electron accelerator ELBE in HZDR (Germany) \cite{10.1103/physrevapplied.13.044054}. The sample was aligned in the ELBE beamline and the beam current was stabilized at $1\,\upmu\mathrm{A}$, which was monitored using a Faraday cup. To achieve uniform defect formation, the sample stage was shifted between four different positions every $5\,\mathrm{h}$, resulting in a total irradiation time of $20\,\mathrm{h}$. This procedure ensured a uniform fluence across the entire 4H-SiC sample.
No post-irradiation annealing were performed. 

\begin{figure}[b]
  \includegraphics[width=.49\textwidth]{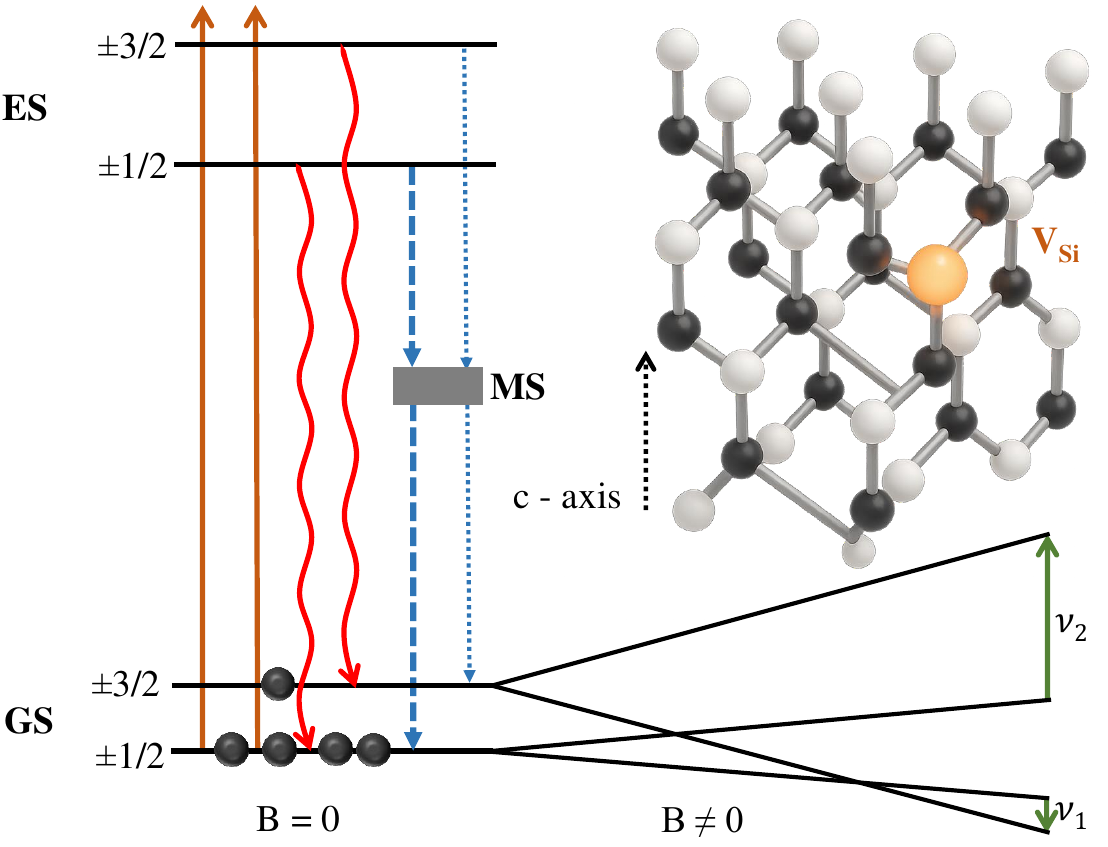}
  \caption{ Energy level diagram of silicon vacancy ($\mathrm{V_{Si}}$): The $\mathrm{V_{Si}}$ spin state $S= 3/2$ is split in two spin sublevels $m_s = \pm3/2$ and $m_s = \pm1/2$ without magnetic field $B = 0$. Optical excitation (solid arrows) are followed by emission back (wavy arrows) from the excited states (ES) to the ground states (GS). The spin dependent relaxation (dashed arrows) through the metastable states (MS) leads to the preferential population of the $m_s = \pm1/2$. In nonzero magnetic fields $B \neq 0$, the spin states are split further, leading to the spin resonances at two frequencies $\nu_1$ and $\nu_2$. The inset shows the 4H-SiC crystal structure with a $\mathrm{V_{Si}}$ defect.}
  \label{figure:figure1}
\end{figure}

We use the V2 center (V$_\mathrm{Si}$(h)), corresponding to a silicon vacancy at the hexagonal site in 4H-SiC, as a room-temperature quantum sensor \cite{10.1038/srep05303}. 
It possesses spin $S = 3/2$ with intrinsic zero-field splitting $2D = 70$~MHz between the $m_s = \pm1/2$ and $m_s = \pm3/2$ spin sublevels \cite{10.1038/nphys2826}, as illustrated in Fig.~\ref{figure:figure1}. Under optical excitation from the ground state (GS) to the excited state (ES), the  subsequent radiative recombination to the GS occurs, which can be detected as PL. In addition to that, non-radiative spin-dependent relaxation through the metastable state (MS) leads to the preferential population of the $m_s = \pm1/2$ GS. The applied MW field can be tuned to the spin resonance transitions of the $\mathrm{V_{Si}}$ defect. Consequently, ODMR spectra are measured as the PL variations depending on the MW frequency. In the absence of an external magnetic field, the $m_s = \pm 1/2$ and $m_s = \pm 3/2$ spin sublevels are degenerate. When a magnetic field ($B \neq 0$) is applied along the $c$-axis, this degeneracy is lifted via the Zeeman effect, resulting in two distinct ODMR transition frequencies, denoted as $\nu_1$ and $\nu_2$ in Fig.~\ref{figure:figure1}.

\begin{figure}[t]
  \includegraphics[width=.49\textwidth]{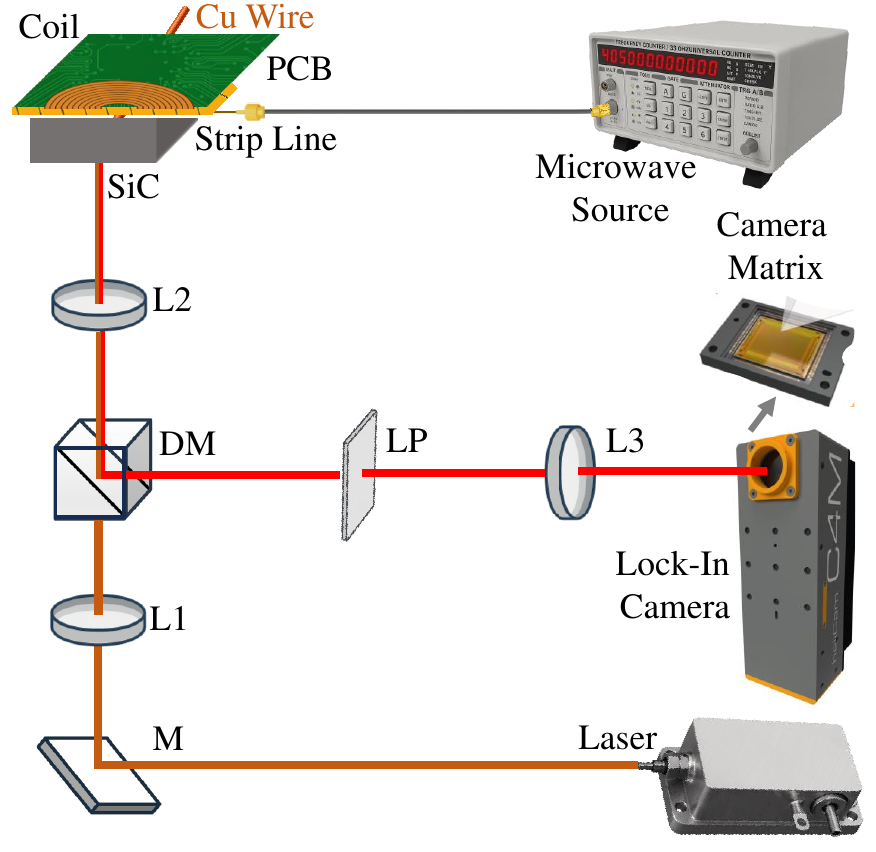}
  \caption{ Schematics of a SiC quantum microscope. A collimated beam from a 795~nm laser is directed via a silver mirror (M) to the lens L1 to focus the beam on the back plane of the imaging objective lense (L2) and form an excitation spot on the 4H-SiC sensor with a diameter of approximately $1 \, \mathrm{m m}$. The PL emitted from the $\mathrm{V_{Si}}$ defects in the 4H-SiC is collected by the same objective lens L2 and spectrally filtered using an 805~nm dichroic mirror (DM) and an 850~nm long-pass filter (LP). The resulting PL is projected by a tube lens (L3) onto a $512 \times 542$ matrix of the lock-in camera. The $\mathrm{V_{Si}}$ spin states in the 4H-SiC sensor are driven by the microwave field $B_1$ delivered via a coaxial cable to a strip line of the coplanar waveguide (CPW) fabricated on a printed circuit board (PCB). A coil positioned above the PCB generates a static bias magnetic field $B_0$ aligned with the $c$-axis of the 4H-SiC crystal. An electrical current passing through a Cu wire creates a spatially varying magnetic field $B_W$, which is imaged via ODMR in 4H-SiC on the camera matrix.}
  \label{figure:figure2}
\end{figure}

A schematic representation of the experimental setup is shown in Fig.~\ref{figure:figure2}. A printed circuit board (PCB) with a waveguide was used to deliver the driving MW field to the $\mathrm{V_{Si}}$ defects in the 4H-SiC sample. A copper (Cu) wire, used for the generation of electrical signals, was placed between the PCB and 4H-SiC sample. To apply a bias magnetic field $B_0$ along the $c$-axis of the crystal, a magnetic coil was positioned above the PCB.

To excite the $\mathrm{V_{Si}}$ defects in the 4H-SiC sample, a $795\,\mathrm{nm}$ laser 
(QPC Lasers) 
was used, operating at a power of $12\,\mathrm{W}$. A silver mirror M 
was used to direct the colimated laser beam towards a lens L1 with a focal length of $100\,\mathrm{mm}$. 
The beam was subsequently focused onto the rear aperture of the objective lens L2 with a focal length of $30\,\mathrm{mm}$.  
The incident beam directed at the 4H-SiC sample had a diameter of approximately $1\,\mathrm{mm}$. PL emitted from the $\mathrm{V_{Si}}$ defects was collected using the same objective lens (L2). A short-pass dichroic mirror DM 
with a cutoff wavelength of $805\,\mathrm{nm}$ was used as a beam splitter and a long-pass filter LP 
with a cut-on wavelength of $850\,\mathrm{nm}$ were used to ensure the PL reaches the lock-in camera. The PL was focused onto a $512 \times 542$ matrix of the lock-in camera using a tube lens L3 with a focal length of $200\,\mathrm{mm}$. 
The lock-in camera was operated in continuous acquisition mode with a modulation frequency of $1\,\mathrm{kHz}$ or $2\,\mathrm{kHz}$. 
The synchronization of the MW modulation and lock-in detection was enabled by the pulse streamer. The software extracts the in-phase ($I = I^+ - I^-$) and quadrature ($Q = Q^+ - Q^-$) signals per pixel, which were integrated over multiple cycles  ($N_{\text{cycle}} ={100}$) and combined into a single frame.

\begin{figure}[t]
  \includegraphics[width=.49\textwidth]{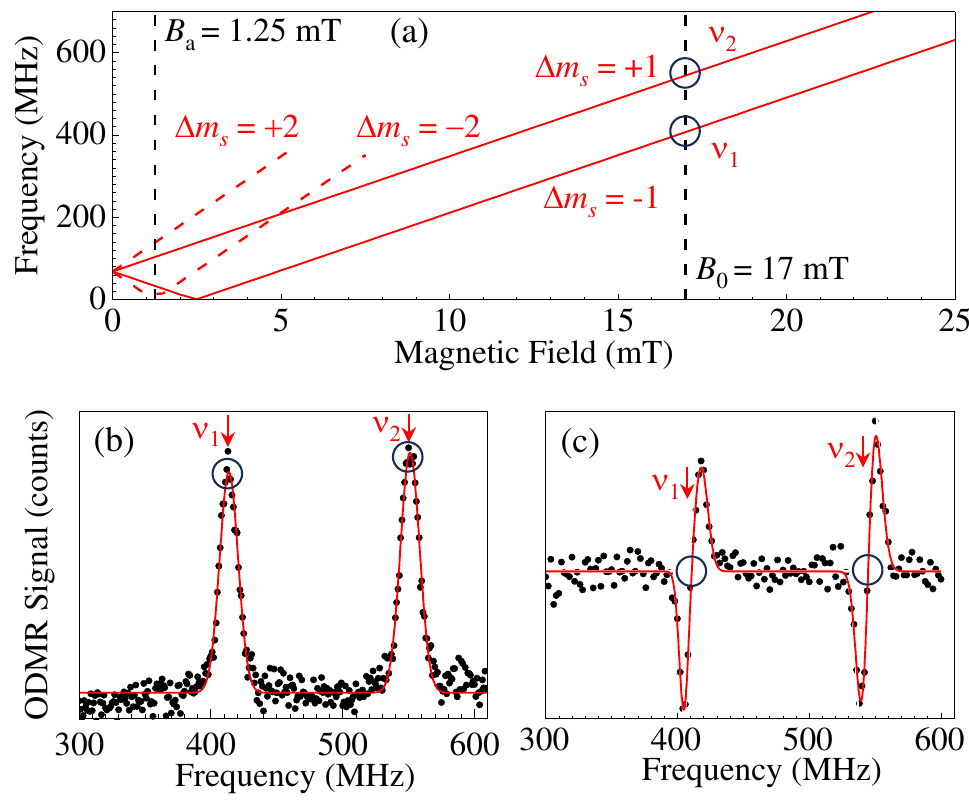}
  \caption{ ODMR spectroscopy in 4H-SiC. (a) The $\mathrm{V_{Si}}$ spin resonance frequencies for the transitions with $\Delta m_s = \pm 1$ (solid red lines) and $\Delta m_s = \pm 2$ (dashed red lines) as a function of the magnetic field applied along the $c$-axis.  The ODMR frequencies $\nu_1$ and $\nu_2$ in the bias magnetic field $B_0 = 17 \, \mathrm{m T}$ are labeled by circles. The magnetic field $B_a = 1.25 \, \mathrm{m T}$ corresponding to the GSLAC-2 is indicated by the vertical dashed line. (b) ODMR spectrum recorded with the lock-in camera using amplitude modulation (AM) at $B_0$, showing ODMR peaks at $\nu_1 = 413 \, \mathrm{MHz}$ and $\nu_2 = 550 \, \mathrm{MHz}$. (c) ODMR spectrum recorded with frequency modulation (FM) at the same $B_0$, revealing a derivative shape. }
  \label{figure:figure3}
\end{figure}

The evolution of the ODMR lines with magnetic field $B$ applied along the $c$-axis is presented in Fig.~\ref{figure:figure3}. The MW driven transitions $\nu_1$ and $\nu_2$ with $\Delta m_s = -1$ and $\Delta m_s = +1$, respectively (solid lines in Fig.~\ref{figure:figure3}) follow the equation \cite{10.1103/physrevapplied.4.014009} 
\begin{equation}\label{ODMR_frequencies}
\nu_{1,2} = | \gamma B \pm 2D | \,.
\end{equation}
Here, $\gamma = 28 \, \mathrm{MHz / mT}$ is the electron gyromagnetic ratio, $2D$ is the zero-field splitting parameter of the $\mathrm{V_{Si}}$ in 4H-SiC and $B$ is the applied magnetic field along the $c$-axis. 

To record ODMR spectra, we use two modulation schemes, namely, amplitude modulation (AM) and frequency modulation (FM). 
In case of AM, we switched the MW field ON and OFF using a square wave modulation signal, and the resulting $I$ phase of the PL signal was demodulated using the lock-in camera, and averaged over a defined region of interest (ROI) on the image sensor. 
The resulted AM ODMR spectrum reveals two peaks shown in Fig.~\ref{figure:figure3}b with spin resonance frequencies $\nu_1 = 413 \, \mathrm{MHz}$ and $\nu_2 = 550 \, \mathrm{MHz}$. 
The bias magnetic field $B_0 = 17 \, \mathrm{mT}$  is obtained from Eq.~(\ref{ODMR_frequencies}) as $B_0 = (\nu_1 + \nu_2) / 2 \gamma$.

In case of FM, the MW frequency was swept with a modulation depth of $\pm 4~\mathrm{MHz}$ using a square wave, and the resulting $I$ phase of the PL signal was demodulated using the lock-in camera and averaged over the ROI. An example of a FM ODMR spectrum is shown in Fig~\ref{figure:figure3}c, exhibits a derivative-like shape that crosses zero at the resonance frequencies $\nu_1$ and $\nu_2$, and was recorded at the same bias magnetic field $B_0$ shown in Fig.~\ref{figure:figure3}a.

\section{\label{sec:results}Results and Discussion}

\subsection{\label{sec:imaging}Imaging Electrical Current}

\begin{figure}[t]
  \includegraphics[width=.49\textwidth]{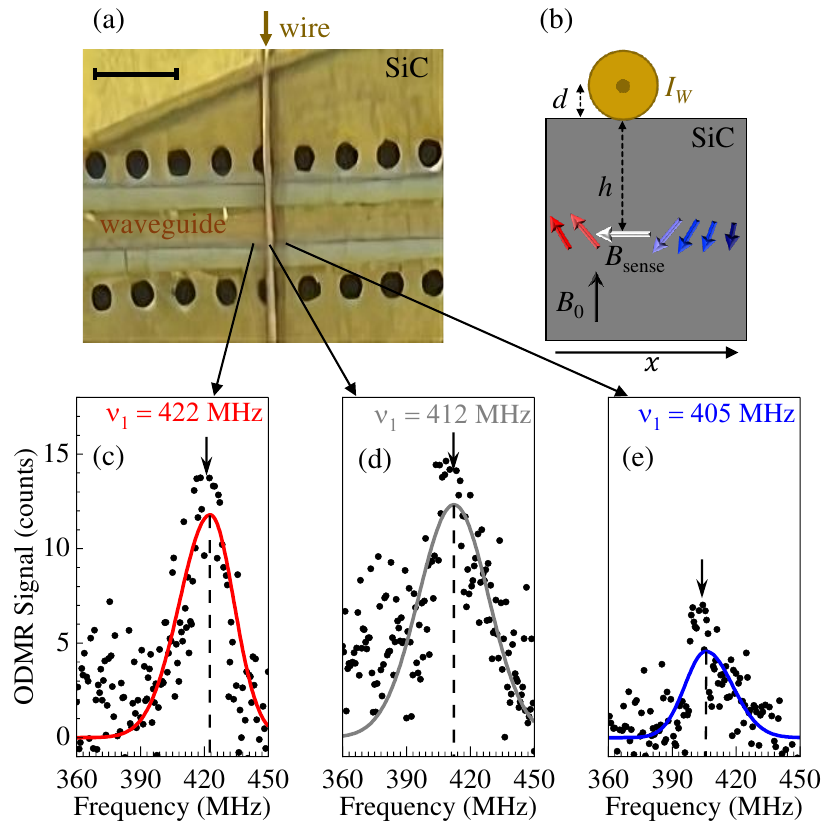}
  \caption{ Spatially resolve detection of magnetic fields generated by electrical currents. (a) A photograph of the Cu wire positioned on the top of 4H-SiC. Electrical current in the wire generates  magnetic fields $B_W$, which lead to the additional spin splitting of the $\mathrm{V_{Si}}$ defects in 4H-SiC with respect to the bias magnetic field $B_0$. This splitting is measured by means of ODMR using MW fields generated by the coplanar waveguide (CPW) below the SiC crystal. The scale bar  is $1 \, \mathrm{m m}$. (b) Side view schematically representing the experimental geometry. The current $I_W = -1 \, \mathrm{A}$ in the  wire with a radius $d$ generates spatially varying magnetic field ($B_W$) inside the 4H-SiC crystal, with the strength and orientation being dependent on the depth below the wire $h$ and the horizontal position $x$ relative to the wire, as schematically depicted by the red, white and blue arrows. The bias magnetic field $B_0$ is applied along the $c$-axis. (c--e) ODMR spectra (dots) obtained from different horizontal positions in the sample relative to the wire using lock-in camera. They show clear frequency shifts of the spin resonance frequency $\nu_1$. From Gaussian fits of the $\nu_1$ ODMR lines (solid lines), we obtain (c) $\nu_1 = 422.0 \pm 0.2~\mathrm{MHz}$ left from the wire, (d) $\nu_1 = 412.2 \pm 0.5~\mathrm{MHz}$ below the wire and (e) $\nu_1 = 405.0 \pm 0.4~\mathrm{MHz}$ right from the wire.}
  \label{figure:figure4}
\end{figure}

A photograph of SiC crystal with a Cu wire is shown in Fig.~\ref{figure:figure4}a. According to the Biot--Savart's law, an electrical current in the wire $I_W$ generates magnetic field around it 
\begin{equation}\label{Total_Bw}
\left| \vec{B}_{W} \right| = \frac{\mu_0 I_{W}}{2\pi R} \,.
\end{equation}
where $\mu_0$ is the permeability of free space, $I_W$ is the current through the wire, and $R = \sqrt{(h + d)^2 + x^2}$ with $d$ is the radius of the wire, $h$ is the depth in the sample, and $x$ is the distance from the wire in the horizontal direction, as depicted in Fig.~\ref{figure:figure4}b. 
Because the $\mathrm{V_{Si}}$ spins are sensitive to the projection of the current-induced magnetic field on the direction of the bias magnetic field $B_0 \parallel c$. Considering the geometry of Fig.~\ref{figure:figure4}b, this sensing magnetic field is   
\begin{equation}\label{Projection_Bs}
B_{\mathrm{sens}} = \frac{\mu_0}{2\pi}  \frac{x}{{(d + h)^2 + x^2}} \,  I_{W} \,.
\end{equation}
where, $I_W$ is the current through the wire. The $\mathrm{V_{Si}}$ defects in SiC respond to $B_{\mathrm{sens}}$,  resulting in spatially-dependent shift of their ODMR frequencies in accord with Eq.~(\ref{ODMR_frequencies}).  

Figures~\ref{figure:figure4}c-e show ODMR spectra for the $\nu_1$ transition left ($x < 0$), below ($x = 0$) and right ($x > 0$) of the wire obtained with the lock-in camera with for the electrical current in the wire $I_W = -1 \, \mathrm{A}$.  
From a Gaussian fit, we obtain $\nu_1 = 422.0 \pm 0.2~\mathrm{MHz}$, $\nu_1 = 412.2 \pm 0.5~\mathrm{MHz}$ and  $\nu_1 = 405.0 \pm 0.4~\mathrm{MHz}$, respectively. These shifts in spin resonance frequency $\nu_1$ reflect the relative change in the magnetic field $B_{\mathrm{sens}}$, thereby validating  Eq.~(\ref{Projection_Bs}) with the experimental detections.

%
%

\begin{figure}[t]
  \includegraphics[width=.49\textwidth]{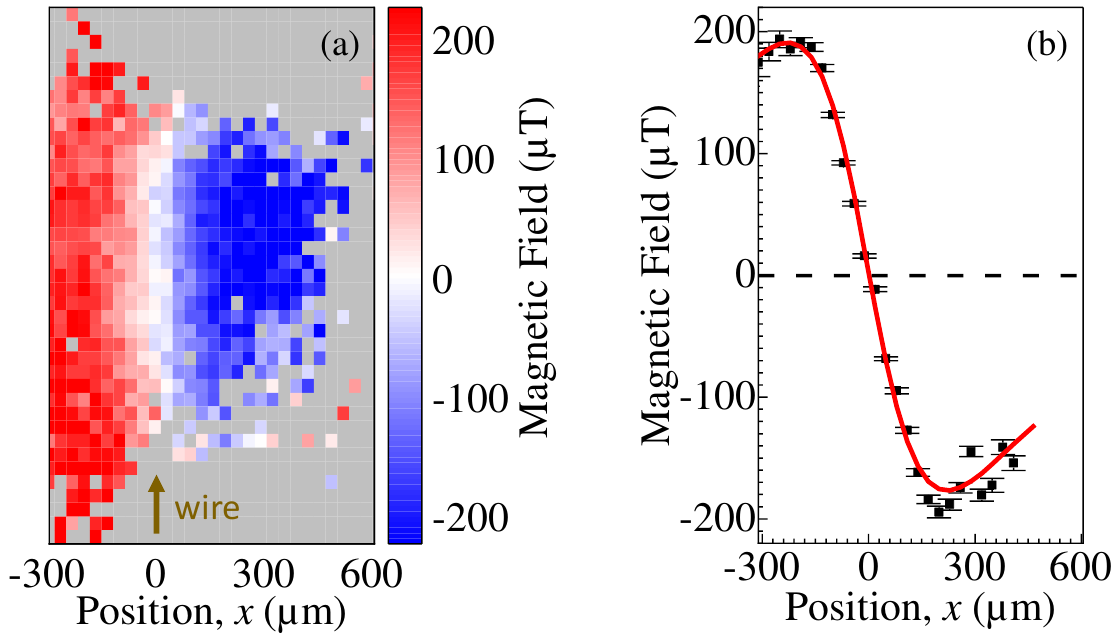}
  \caption{  Magnetic field  mapping. (a) Magnetic field map reconstructed from the ODMR spectra recorded with a lock-in camera using $10 \times 10$ pixel binning. The color code indicates the magnetic field component generated by the current $I_W = 1 \, \mathrm{A}$ in the wire, with the zero-field region located directly beneath the wire. Gray regions denote pixels excluded by a threshold-based filter, as described in the algorithm presented in the text. (b) Symbols show the linear profile of the magnetic field component $B_W$ across the wire, extracted from the magnetic field map in (a). The solid line shows the calculation with the current $I_W = -1 \, \mathrm{A}$. }
  \label{figure:figure5}
\end{figure}

The magnetic field map $B_{\mathrm{sens}}$ generated by the current in the Cu wire is shown in Fig.~\ref{figure:figure5}a. A binning of $10 \times 10$ pixel is applied on the raw image to enhance the sensitivity. A custom Python-based software was used to analyze the data, where the $\nu_1$ frequency was obtained from automatic Gaussian fit for each binned pixel. A software-based threshold filter was applied to eliminate bad pixels with noisy signal, specifically those located away from the waveguide where the ODMR signal is not detectable. These excluded pixels are shown in the gray region of Fig.~\ref{figure:figure5}a. The magnetic field strengths for good pixels are then obtained as $B_{\mathrm{sens}} = B - B_0$, where $B$ is calculated from $\nu_1$ using Eq.~(\ref{ODMR_frequencies}).   

The reconstructed 2D map of Fig.~\ref{figure:figure5}a shows the magnetic field distribution across the Cu wire, indicating the field variation from $B_{\mathrm{sens}} = +200\, \mathrm{\mu T}$ to $B_{\mathrm{sens}} = -200\,\mathrm{\mu T}$. The region directly under the wire $x = 0$ is imprinted in the map as $B_{\mathrm{sens}} = 0  \, \mathrm{\mu T}$, line as expected from Eq.~(\ref{Projection_Bs}).  The linear profile in Fig.~\ref{figure:figure5}(b) presents the one-dimensional variation of $B_{\mathrm{sens}}$, obtained by averaging the 2D magnetic field map along the $y$ direction, i.e., parallel to the wire, and plotting the resulting line profile as a function of the $x$ position. 

The solid line in Fig.~\ref{figure:figure5}(b) is a fit using Eq.~(\ref{Projection_Bs}) with in-depth integration from $h =0 \, \mathrm{\mu m}$ to $h = 300 \, \mathrm{\mu m}$ to account for the sample thickness. From the best fit, we found  $d = 220 \pm 4\, \mathrm{\mu m}$, corresponding to the wire radius and the distance from the wire to the SiC surface,  and $I = - 1.00 \pm 0.04\, \mathrm{A}$.
The agreement between the measured and calculated values confirms the accuracy of the imaging method. It facilitates spatial localization of the wire in the absence of a direct line of sight and enables current measurement without electrical contacts.

\subsection{\label{sec:dual-frequency} Dual-frequency imaging of electrical pulses}

\begin{figure}[t]
  \includegraphics[width=.49\textwidth]{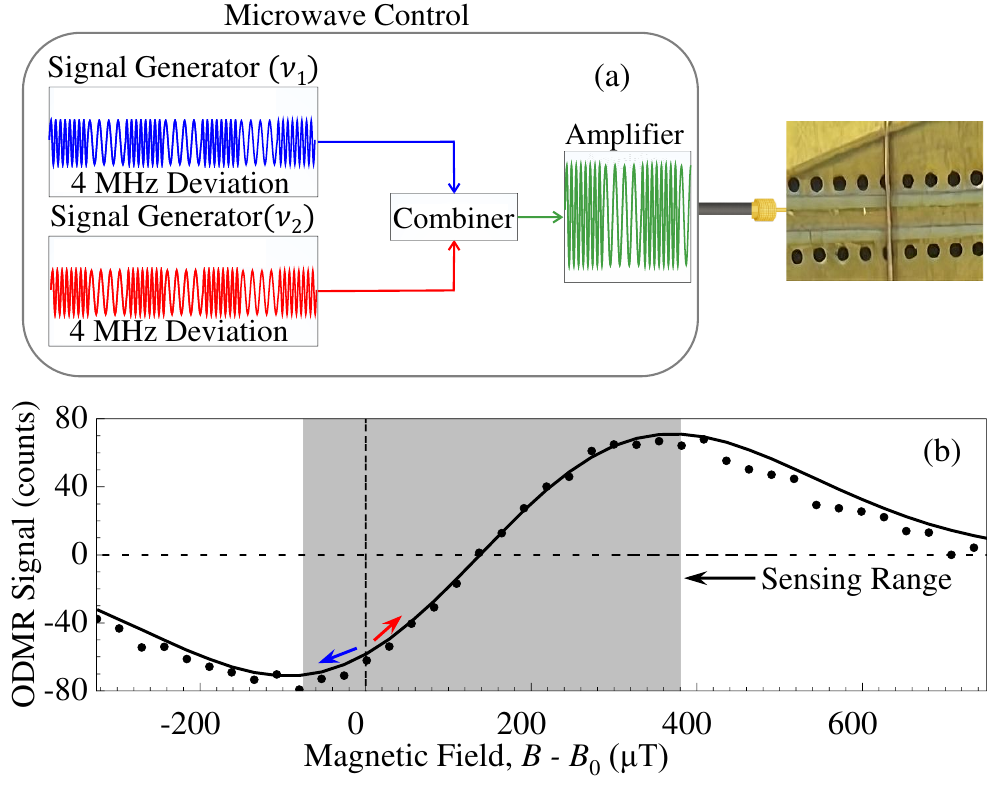}
  \caption{ Dual-frequency sensing using FM ODMR.  (a) Schematics of the microwave control with two signal generators working at the ODMR resonance frequencies $\nu_1$ and $\nu_2$ 
    modulated with a 4\,MHz frequency deviation. The microwave fields are combined together and amplified before transmitted through the coaxial cable to the CWG. (b) Calibration curve of our SiC quantum microscope obtained from a small variation of the magnetic field with respect to the bias magnetic field $B - B_0$. The black dots represent the experimental data, and the solid line is a derivative-Gaussian fit. The sensing range is represented by the shaded area.}
    \label{figure:figure6}
\end{figure}

The dual-frequency technique is used to improve  magnetic field sensitivity by simultaneously driving the two spin transitions with $\Delta m_s = -1$ and $\Delta m_s = +1$. 
The MW fields from two signal generators at $\nu_1 = 368\,\mathrm{MHz}$ and $\nu_2 = 493\,\mathrm{MHz}$ shown in Fig.~\ref{figure:figure6}a were frequency modulated, 
then combined and amplified before being sent to the 4H-SiC sample via the coplanar waveguide (CWG). The simultaneous application of two frequencies enhances the readout contrast and reduces noise caused from strain and temperature fluctuations, which lead to variations in the zero-field splitting parameter $D$ \cite{10.1103/physrevapplied.19.044086, 10.1038/s41534-025-01011-2}. Since both spin transitions $\Delta m_s = -1$ and $\Delta m_s = +1$ are affected similarly by common disturbances, their signals cancel out the common noise while doubling the contribution from the magnetic field, which can be determined using Eq.~(\ref{ODMR_frequencies}). This enables the detection of weak magnetic fields produced by small currents with temporal resolution through dual-frequency sensing. 

The corresponding calibrated sensitivity curve 
is presented in Fig.~\ref{figure:figure6}b, which demonstrates how the ODMR signal responds to change in the bias magnetic field $B - B_0$. It exhibits a derivative-like shape due to FM, where the zero-crossing corresponds to the resonance frequency and is highly sensitive to small changes in the magnetic field induced by the applied current. The shaded region in Fig.~\ref{figure:figure6}b highlights the sensing range ($-75\,\upmu\mathrm{T}$ to $+380\,\upmu\mathrm{T}$), where each measured signal value uniquely corresponds to a specific magnetic field magnitude. This range is shifted with respect to $B_0$ due to the slight detuning of the driving MW frequencies $\nu_1$ and $\nu_2$ from the resonance spin transitions $\Delta m_s = -1$ and $\Delta m_s = +1$. This detuning originates from slow temporal fluctuations of the environmental magnetic field, which define the effective background field $B_0$.

\begin{figure}[t]
  \includegraphics[width=.49\textwidth]{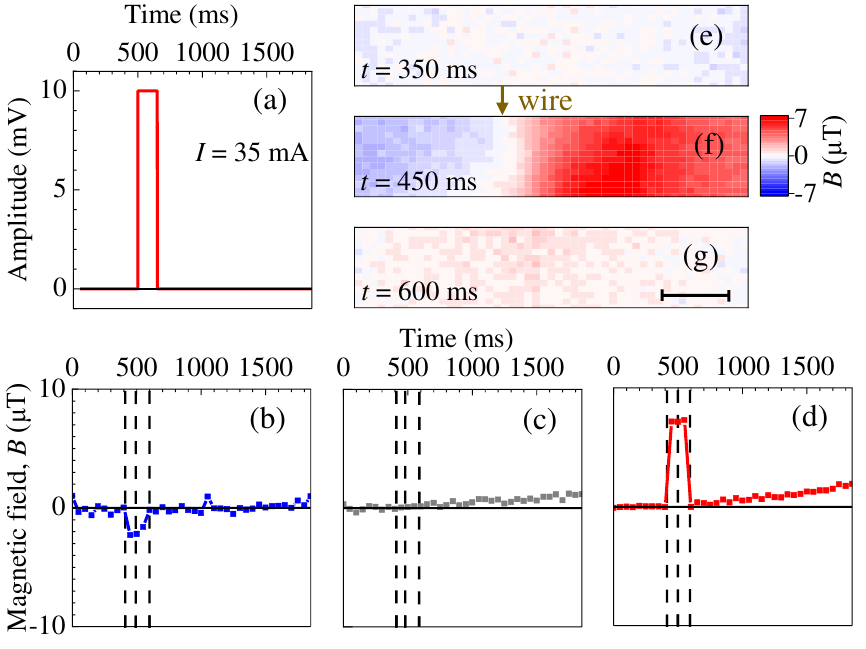}
  \caption{  Imaging of electrical signals. (a) Voltage pulse  applied to a Cu wire with a duration of $100 \,\mathrm{ms}$ generates an electrical current pulse with an amplitude of $35 \,\mathrm{mA}$. (b) Magnetic field pulse measured on the left side of the wire. (c) Magnetic field is almost zero, shows the region below the wire. (d) Magnetic field pulse measured on the right side of the wire. (e-g) Magnetic field maps at different time intervals:  at $t = 350 \,\mathrm{ms}$, i.e., before the pulse (e), at $t = 450 \,\mathrm{ms}$, i.e., during the pulse (f), and at $t = 600 \,\mathrm{ms}$, i.e., after the pulse (g). The scale bar is  $200~\mu\mathrm{m}$.}
  \label{figure:figure7}
\end{figure}

The dual-frequency ODMR sensing was used to detect dynamic magnetic fields and capture time-resolved imaging of pulsed electric currents. A voltage pulse of $10\,\mathrm{mV}$ from pulse generator with a duration of $100\,\mathrm{ms}$ (Fig.~\ref{figure:figure7}a) was applied to the Cu wire positioned under the 4H-SiC sample to image the electrical signals. This generated a current pulse, which, in turn, induces a spatially and temporally varying magnetic field. The magnetic field response was reconstructed from the ODMR signal counts by applying the calibrated sensitivity curve illustrated in Fig.~\ref{figure:figure6}b. 
Time-resolved magnetic field images were recorded using the lock-in camera, which was synchronized via a TTL trigger from the pulse generator. 
A controlled delay of $500\,\mathrm{ms}$ was introduced between the voltage pulse and the start of acquisition. Each sequence consisted of 40 frames over $2\,\mathrm{s}$, and to improve the signal-to-noise ratio, the sequence was repeated $100$ times.


Figures~\ref{figure:figure7}b–d show the temporal evolution of the magnetic field changes in a single $10 \times 10 $ binned pixel over time on the left, below and right sides of the wire. A clear magnetic field signal was observed on both sides of the wire, as shown in Fig.~\ref{figure:figure7}b and Fig.~\ref{figure:figure7}d. However, the amplitude of the magnetic field observed on the left side in Fig.~\ref{figure:figure7}b is smaller than that shown in Fig.~\ref{figure:figure7}d. This difference arises because the negative magnetic fields generated by the current in the Cu wire fall outside the calibration range used to convert the lock-in signal to magnetic field strength (see Fig.~\ref{figure:figure6}b). In contrast, the positive field on the right side remains within the calibrated range, resulting in a more accurate reconstruction of the magnetic field amplitude. The field below the wire remained close to zero, as illustrated in Fig.~\ref{figure:figure7}c, which is consistent with the expected spatial symmetry of the generated magnetic field.

After the pulse, a slight increase in the signal is observed in Fig.~\ref{figure:figure7}d. This is likely caused by a small mismatch between the modulation frequency used in the dual-frequency microwave technique and the demodulation frequency of the lock-in camera. As a result, a phase shift accumulates over time after the trigger, particularly when the magnetic field disappears. Since the magnetic field signal is extracted as the in-phase component, $I = I^+ - I^-$, this mismatch prevents the perfect cancellation of $I^+$ and $I^-$ in the the frames acquired at longer times after the trigger. This effect leads to the slight upward drift in the baseline after the current pulse for $t > 700\,\mathrm{ms}$. 

In this experiment, the electric current associated with the voltage pulse was not measured directly. Instead, it was inferred by comparing the measured magnetic field profile shown in Fig.~\ref{figure:figure5}b. This calibration curve, obtained for a known current using Eq.~(\ref{Projection_Bs}), provides a direct relationship between the magnetic field and the applied current in the Cu wire. Based on this calibration, the applied electric current during the voltage pulse was found to be $35\,\mathrm{mA}$, suggesting the sensitivity increased by an order of magnitude and the ability to measure electrical current without contacts. 

To analyse the time-resolved characteristics of the signal, magnetic field maps were extracted at three time intervals: before the current pulse at $t = 350\,\mathrm{ms}$, during the pulse at $t = 450\,\mathrm{ms}$, and after the pulse at $t = 600\,\mathrm{ms}$ (Fig.~\ref{figure:figure7}e--g). No magnetic field is observed before the pulse (Fig.~\ref{figure:figure7}e). During the pulse, a clearly symmetric magnetic field pattern appears (Fig.~\ref{figure:figure7}f), with the zero-crossing point in the map corresponding to the wire position. After the pulse ends, the magnetic field disappears again (Fig.~\ref{figure:figure7}g). This confirms the capability for time-resolved imaging of transient electrical signals.

\subsection{\label{sec:MW-free} Microwave-free quantum imaging}

\begin{figure}[t]
   \includegraphics[width=.48\textwidth]{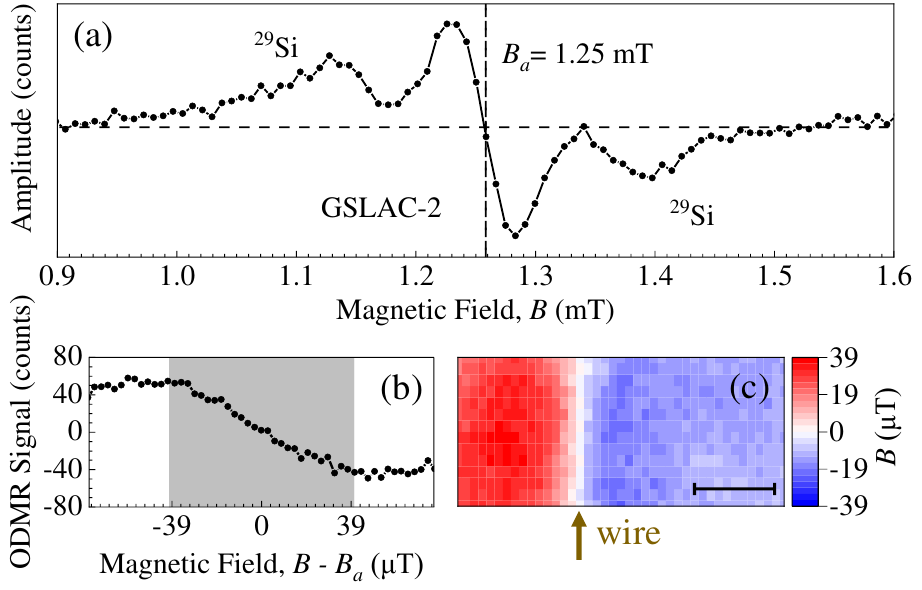}
  \caption{ Microwave-free magnetic field imaging. (a) PL variation in the external magnetic field $B$ in the vicinity of GSLAC-2 at $B_a = 1.25\,\mathrm{mT}$. The satellites are due the hyperfine interaction with \textsuperscript{29}Si nuclei. (b) Calibration curve for microwave-free sensing obtained from a small variation of the magnetic field with respect to the GSLAC-2 field $B - B_a$. The sensing range from $-39~\mu\mathrm{T}$ to $+39~\mu\mathrm{T}$ is represented by the shaded area. (c) Magnetic field map reconstructed from the calibration curve (b), showing the field generated by the current-carrying wire. The scale bar is $200~\mu\mathrm{m}$.}
  \label{figure:figure8}
\end{figure}

Recent advances in quantum sensing have highlighted the promise of MW-free protocols that exploit the energy level anticrossings of solid-state spin defects at specific magnetic fields \cite{10.1063/1.4960171, 10.1103/physrevx.6.031014, 10.1103/PhysRevApplied.21.044039}. This approach, known as MW-free quantum imaging, is attractive because it eliminates the need for MW fields and simplifies technical requirements. Achieving a uniform MW field for widefield imaging requires high MW power, which can be challenging to implement. In some applications, like with liquid or complex multilayer electronic circuits, MWs may cause heating or be reflected or screened. This makes difficult to record  ODMR spectra.

In our experiment, the MW-free magnetic field imaging was performed using the ground-state level anti-crossing (GSLAC) in V2 (V$_\mathrm{Si}$(h)) centers in 4H-SiC. A prominent change in PL occurs at an external magnetic field $B_a = 1.25\,\mathrm{mT}$, corresponding to GSLAC-$2$, where the spin sublevels $m_S = -3/2$ and $m_S = +1/2$ undergo mixing (Fig.~\ref{figure:figure1} and Fig.~\ref{figure:figure3}a). At this anti-crossing point, the spin populations are redistributed between the $m_S = -3/2$ and $m_S = +1/2$ levels due to their nearly identical energies and mixing interactions, leading to a measurable change in PL intensity. By sweeping the external magnetic field around $B_a$, the mixed state changes the PL without the need for any MW field. To detect the small magnetic field signals induced by the current, magnetic field modulation with a modulation depth of approximately $\pm 50\,\upmu\mathrm{T}$ and modulation frequency of $2\,\mathrm{kHz}$  was used instead of MW  modulation, and the lock-in camera was used to record the PL signal. 

Figure~\ref{figure:figure8}a shows the PL change with the external magnetic field near the GSLAC-2 point at $B_a = 1.25\,\mathrm{mT}$, where the $m_S = -3/2$ and $m_S = +1/2$ spin sublevels mix. The additional smaller peaks that appear on either side of the main GSLAC-2 arise from hyperfine interactions with ${}^{29}\mathrm{Si}$ nuclei \cite{10.1103/physrevx.6.031014}. The resulting calibration curve for the ODMR signal as a function of $B - B_a$ is presented in Fig.~\ref{figure:figure8}b, with a sensing range from $\delta_- = -39\,\upmu\mathrm{T}$ to $\delta_+ =  +39\,\upmu\mathrm{T}$. It corresponds to the sensitivity for an effective ODMR linewidth of $\gamma (\delta_+ - \delta_-) = 2.2,\mathrm{MHz}$, representing an improvement by one order of magnitude compared single-frequency ODMR in Fig.~\ref{figure:figure4}, as explained earlier  \cite{10.1103/physrevx.6.031014}. 

The resulted spatial field map reconstructed from this calibration is shown in Fig.~\ref{figure:figure8}c. 
The field distribution around the Cu wire shows a change in sign across the wire, ranging from a sensing range of $\delta_- = -39\,\upmu\mathrm{T}$ to $\delta_+ = +39\,\upmu\mathrm{T}$. The zero-field line indicates the position directly beneath the wire at $x = 0$, aligning with the behavior seen in the ODMR-based magnetic field maps of electrical current imaging and dual-frequency sensing in Fig.~\ref{figure:figure5}a and Fig.~\ref{figure:figure7}e–g. This verifies that the MW-free sensing protocol consistently reproduces the spatial profiles of the magnetic field. By using the calibrated sensitivity curve from Fig.~\ref{figure:figure5}b, the current in the reconstructed magnetic field map was found to be $100\,\mathrm{mA}$, which aligns exactly with the applied current of $100\,\mathrm{mA}$. This confirms the accuracy and reliability of the microwave-free sensing protocol. This method uses the GSLAC-2 feature of $\mathrm{V_{Si}}$ centers to enable MW-free sensing, making it suitable for environments where MW application is challenging.

\section{Conclusion}

We developed and experimentally tested a quantum silicon carbide microscope capable of widefield magnetic imaging using both dual-frequency ODMR and microwave-free protocols. Using the quantum properties of $\mathrm{V_{Si}}$ spin defects in 4H-SiC, we successfully imaged electrical currents and reconstructed the corresponding magnetic fields with $50 \times 50$ virtual pixels, $30\,\mathrm{\mu m}$ spatial resolution, and $50\,\mathrm{ms}$ temporal resolution. 
Furthermore, we developed a custom Python software to process and visualize the images. This enables wide field magnetic imaging  in a single frame, along with time-resolved characterization of the magnetic field evolution in response to the applied current pulse. The MW-free method based on the spin level anticrossing further extended the system's relevance to environments in which MW application is technically challenging.

To estimate the magnetic field sensitivity, we use the noise floor of about $1 \, \mathrm{\mu T}$ in  Fig.~\ref{figure:figure7}b-d. 
With integration time per cycle of $50 \, \mathrm{ms}$ and $100$ sequences, it gives the sensitivity of  $2.2\,\mathrm{\mu T \, Hz^{-1/2}}$ per virtual pixel with $10 \times 10$ binning at a modulation frequency of $2\,\mathrm{kHz}$. Since we use a c-plane HPSI SiC wafer, where the c-axis is perpendicular to the sample surface used for sensing, the excitation and collection efficiency of the PL from the V2 (V$_\mathrm{Si}$(h)) defects, which emit predominantly perpendicular to the c-axis, is significantly limited \cite{10.1016/j.physb.2009.09.023, 10.1038/s41534-022-00534-2}.  A substantial enhancement in sensitivity is anticipated by employing an a-plane SiC wafer, although such wafers are less commonly available commercially, in combination with a light-trapping waveguide  \cite{10.1038/nphys3291}. Further orders-of-magnitude improvement, potentially reaching sensitivities on the order of $1 \, \mathrm{nT \, Hz^{-1/2}}$, can be achieved by employing epitaxial, isotopically purified $\mathrm{^{28}Si^{12}C}$ layers with thicknesses of tens to hundreds of $\mathrm{\mu m}$ \cite{10.1103/prxquantum.3.010343, 10.1103/physrevapplied.19.044086}. This approach enhances crystalline quality and reduces residual strain, both of which are critical for high-performance quantum sensing. The recently demonstrated protocols for multiplexed QDM with highly-sensitive EMCCD camera with the ability to detect optically single spins  can be applied to QSICM as well \cite{10.1103/t8fz-3tzs, 10.1103/jdzq-jbfz}.  

In summary, our results highlight the strong potential of the QSiCM for advanced sensing and imaging applications.

\section*{Acknowledgement}

This work was supported via project DiaQNOS (contract number 13N16461) from  German Federal Ministry of Research, Technology and Space (BMFTR) and project Quantum Sensing for Fundamental Physics (QS4Physics) from the Innovation pool of the research field Helmholtz Matter of the Helmholtz Association. Partial support from the ELBE and  IBC facilities at the HZDR, a member of the Helmholtz Association, is also acknowledged.

\bibliographystyle{apsrev4-2}
\bibliography{Imaging}

\end{document}